\documentstyle[12pt]{article}
\input psfig.tex
\begin{document}

\begin{titlepage}

\title{\bf {The Force Exerted by a Molecular Motor}}
\author{Michael E. Fisher \  and Anatoly B. Kolomeisky \\
\\
Institute for Physical Science and Technology,\\
University of Maryland, College Park, \ MD 20742 USA}
\date{}

\maketitle
\thispagestyle{empty}

\begin{abstract}

{The stochastic driving force exerted by a single molecular motor (e.g., a kinesin, or myosin) moving on a periodic molecular track (microtubule, actin filament, etc.) is discussed from a general viewpoint open to experimental test.  An elementary ``barometric" relation for the driving force is introduced that (i) applies to a range of kinetic and stochastic models, (ii) is consistent with more elaborate expressions entailing explicit representations of externally applied loads and, (iii) sufficiently close to thermal equilibrium, satisfies an Einstein-type relation in terms of the velocity and diffusion coefficient of the (load-free) motor.   Even in the simplest two-state models, the velocity-vs.-load plots  exhibit a variety of contrasting shapes (including nonmonotonic behavior). Previously suggested bounds on the driving force are shown to be inapplicable in general by analyzing discrete jump models with waiting time distributions. } 

\noindent \rule{132mm}{0.02in}

\end{abstract}

\end{titlepage}

\section*{ }

Molecular motors are protein molecules such as myosin, kinesin,  dynein, and RNA polymerase, that move  along linear tracks  (actin filaments, microtubules, DNA) and perform tasks vital to the life of the organism ---  muscle contraction, cell division, intracellular transport, and genomic transcription (1-5).  Understanding how they operate represents a significant challenge.  The hydrolysis of adenosine triphosphate (ATP), with the release of adenosine diphosphate (ADP) and inorganic phosphate (P$_{\rm i}$), is known to be the power source for many  motor proteins.  An activated motor may well be in a dynamical or, better, a stochastic steady state but it {\it cannot} be in full thermal equilibrium. 

Striking {\it in vitro} experiments observing individual motor proteins moving under controlled external loads  (6\hspace{0.4mm}-11) have stimulated enhanced theoretical work aimed at understanding the mechanisms by which a biological motor functions.  From a broad theoretical perspective, a  molecular motor is a microscopic object that moves predominantly in one direction  along a ``polarized" one-dimensional periodic structure, namely, the molecular track (1-11). In recent years, in addition to traditional chemical kinetic descriptions (see, e.g., (12) and references therein)  and various more detailed schemes  (11,13,14), so-called ``thermal ratchet" models have been proposed to account for the mechanics: see the review (15).

A common feature of most approaches is that a motor protein molecule is associated with a labeled site $l$ ($=0, \pm 1, \pm 2$,  $\ldots$) on the track and is pictured as being in one of $N$ essentially discrete states $j$, which may be {\it free of} (say, $j=0$), or {\it bound to} ATP and its various hydrolysis products ($j=1,2,\ldots,N-1$). Thus, for a kinesin molecule, K, on a microtubule, M, the ($N=4$) states identified might be  M$\cdot$K,  M$\cdot$K$\cdot$ATP,  M$\cdot$K$\cdot$ADP$\cdot$P$_{\rm i}$, and  M$\cdot$K$\cdot$ADP (8,12). Transition rates between these states  can be introduced via
\begin{equation}
\begin{array}{ccccccccc}
     &u_{1}             &      &u_{2}             &        &u_{N-1}           &       &u_{N}             &      \\
0_{l}&\rightleftharpoons&1_{l} &\rightleftharpoons& \cdots &\rightleftharpoons&(N-1)_{l}&\rightleftharpoons&0_{l+1},  \\
     &w_{1}             &      &w_{2}             &        &w_{N-1}           &       &w_{N}            &         
\end{array}
\end{equation}
where the subscripts indicate that the states $j$ are associated with successive sites, $l$ and $l+1$, on the track spaced at distances $\Delta x=x_{l+1}-x_{l}=d$ : this defines the {\it step size} $d$. Of course, states $j_{l}, \ j_{l+1}, \ldots, \ j_{l+n}$ differ physically only in their spatial displacements $d,2d, \ldots, nd$, along the track. By the same token, the rates $u_{j}$ and $w_{j}$ are independent of $l$ (or $x=ld$); however, in the subsequent developments it proves useful to allow for spatially dependent rates  $u_{j}(l)$ and $w_{j}(l)$.

To properly represent physicochemical reality (that is, microscopic reversibility) {\it none} of the forward rates, $u_{j}$, or backward rates $w_{j}$ may strictly vanish even though some, such as the last reverse rate, $w_{N}$, might be extremely small (11,12). On the other hand, if, as one observes in the presence of free ATP, the motor moves under no external load to the right (increasing $x$), the transition rates {\it cannot} (all) satisfy the usual conditions of detailed balance that would  characterize thermal equilibrium if Eq. {\bf 1} were regarded as a set of chemical reactions (near equilibrium) between effective species $j_{l}$ (15). (Notice that one may envisage a second-order rate process, e.g.,  M$\cdot$K + ATP $\rightleftharpoons$ M$\cdot$K$\cdot$ATP, to conclude  $u_{1}=k_{1}$[ATP]; this can then lead to Michaelis-Menten type rate-vs.-concentration relations (6).  However, one might also contemplate a small ``spontaneous" or first-order background rate, $u_{1,0} >0$, that exists even in the absence of ATP.)

Now, within statistical physics, the kinetic scheme in \linebreak Eq. ${\bf 1}$ represents a one-dimensional hopping process of a particle on a periodic but, in general, {\it asymmetric} lattice. After initial transients, the particle will move (16) with steady (mean) velocity $V$, and diffuse (with respect to the mean position, $\overline{x}=Vt$, at time $t$) with a diffusion constant $D$. Complicated, but exact equations for $V$ and $D$ in terms of $u_{J}$ and $w_{J}$ have been obtained for all $N$ (16), as exhibited in the Appendix. A dimensionless, overall rate factor which, rather naturally, appears (see  Eq. ${\bf A1}$), is given by the product
\begin{equation}
\Gamma=\prod_{j=0}^{N-1} \left (\frac{u_{j}}{w_{j}} \right) \equiv e^{\mbox{\scriptsize{\boldmath $\varepsilon$}}}.
\end{equation}
This will play an important role in our discussion. Note, indeed, that viewing Eq. {\bf 1} as a standard set of chemical reactions and requiring  detailed balance would impose $\Gamma \equiv 1$ (or ${\mbox{\boldmath $\varepsilon$}}=0$) whereas $\Gamma >1$ (or ${\mbox{\boldmath $\varepsilon$}} > 0$) is needed for a positive velocity $V$. (One might comment, however, (17) that as regards the full chemistry, the complex of motor protein plus track may be regarded simply as catalyzing the hydrolysis of ATP: the reaction rates for this overall process may then be expected to satisfy detailed balance.)

The simplest or ``minimal" physical models have $N=2$, and one can then calculate analytically not only the steady state behavior but also the full transient responses, specifically, the probabilities, $P_{j}(l;t)$, of being  in state $j_{l}$ (``at" site $l$) at time $t$ having started, say, at site $l=0$ in state $j=0$ at time $t=0$: see (17).  In Ref. (17) only the special (limiting) cases with $w_{N} \equiv w_{2}=0$ were treated; but as seen below, this limit can be misleading and so  for completeness (and for possible comparisons with experiment and simulation), we present the general $N=2$ results: see Appendix. In particular, the velocity and diffusion constant for $N=2$ are 
\begin{equation}
V=\frac{(u_{1}u_{2}-w_{1}w_{2})d}{u_{1}+u_{2}+w_{1}+w_{2}} \equiv (\Gamma-1) \omega d,
\end{equation}
\begin{equation}
D={\textstyle \frac{ 1}{ 2}} \left[ \Gamma+1-2(\Gamma-1)^{2} \omega/\sigma \right] \omega d^{2},
\end{equation}
where $\Gamma=u_{1}u_{2}/w_{1}w_{2}$ (as in  Eq. ${\bf 2}$) and, for convenience, we have introduced the associated overall rates 
\begin{equation}
\sigma=u_{1}+u_{2}+w_{1}+w_{2}, \hspace{8mm} \omega=w_{1}w_{2}/\sigma.
\end{equation} \renewcommand{\thefootnote}{\fnsymbol{footnote}}

One can, of course, envisage more complicated schemes than  Eq. ${\bf 1}$, with various internal loops, parallel pathways, etc.\footnote{ Thus a backwards reaction directly from, say, state $j_{l}^{\dagger}$ to $0_{l}$ could account for ``futile'' ATP hydrolysis, i.e., without forward motion of the motor (13); but note that within $N=2$ models (which enforce $j^{\dagger}=1$) the phenomenon may be described simply by including the futile-hydrolysis  parallel reaction rate in the backward rate $w_{1}$.}    In all cases, however, there will be a well defined (zero-load) steady-state velocity $V$ and a diffusion constant $D$ (independent of the particular states, $j$) which are susceptible to estimation by simulation even though their explicit mathematical expressions may be intractable.  Furthermore, in real systems both $V$, as often demonstrated (6,\hspace{0.4mm}8\hspace{0.4mm}-11), and $D$ (7) are susceptible to experimental measurement.

Now there arises an obvious but crucial question, namely: ``What (mean) driving force, $f$, will such a general motor protein model exert as it moves along the track?" That is the issue we address here from a theoretical standpoint.

\section*{\hspace{5cm} \bf ANALYSIS}

\hspace{0.9em} {\bf Maximum Driving Force.} \hspace{0.6em} The hydrolysis of one ATP molecule releases a free energy $\Delta G_{0}$ of about $0.50 \times 10^{-19}$J (corresponding to $7.3$ Kcal/M or $12  \ k_{B}T$ at typical {\it in vitro} temperatures, $T$ (3)).  If all this free energy could be converted into mechanical energy and move the motor protein through a distance $\Delta x=d$, the step size (going from state $0_{l}$ to $0_{l+1}$: see  Eq. ${\bf 1}$), the force exerted would be
\begin{equation}
f_{ max}=\Delta G_{0}/d.
\end{equation}
Accepting that one molecule of ATP is sufficient to translocate the motor protein by one step (7), this expression clearly represents the maximal driving force that can be exerted. For a kinesin moving on a microtubule (6\hspace{0.4mm}-12) with $d \simeq 8.2 $ nm (9) it yields $f_{max} \simeq 6.2$ pN. If $f$ is the driving force actually realized, the efficiency of a motor protein may sensibly be defined by $\mbox{\LARGE${\varepsilon}$}=f/f_{max}$.

{\bf Einstein Force Scale.} \hspace{0.3em} Consider a small (``mesoscopic") particle with ``instantaneous" position $x(t)$ and velocity $v(t)$ that undergoes one-dimensional Brownian motion in a fixed, slowly varying external potential, $\Phi(x)$. Under a constant external force, $F=-(d\Phi/dx)$, the particle diffuses with a diffusion constant which, for long times, $t$, satisfies
\begin{equation}
D \approx \left[ \langle x^2(t) \rangle- \langle x(t) \rangle^{2} \right] /2t,
\end{equation}
where $\langle \hspace{2mm} \cdot \hspace{2mm} \rangle$ denotes a statistical average  (18\hspace{0.4mm}-21). In addition, the particle experiences an (effective) {\it frictional force}, $f_{E}=\zeta v(t)$, where $\zeta$ is a friction coefficient determined by the environment (18\hspace{0.4mm}-21). In a steady state, the friction balances the external force, $F$, leading to a drift motion, $\langle x(t) \rangle \approx Vt$, with mean velocity $V=F/\zeta=f_{E}/\zeta$.  Now, by definition, Brownian motion takes place within {\it full thermal equilibrium}: that fact dictates  (18\hspace{0.4mm}-21) the Einstein relation $\zeta=k_{B}T/D$ which, in turn, implies the result
\begin{equation}
f_{E}=k_{B}TV/D.
\end{equation}

In the present context this is an appealing formula in that it sets a {\it force scale} in terms only of the velocity, $V$, and the diffusion constant, $D$, predicted by a motor-protein model (or observed in an experiment or simulation); one might call it the ``Einstein scale". However, because an activated molecular motor is {\it not} a Brownian particle and {\it cannot} be described by thermal equilibrium, there are no grounds for expecting $f_{E}$ to be related to the proper driving force, $f$. Nevertheless, we will show that in a certain limit such a Brownian motion ``mimic" of an activated motor protein does provide an appropriate prediction for $f$.  Indeed, Ref. {17} accepted the identification $f=f_{E}$ without discussion and used Eq. {\bf 8} to estimate driving forces for restricted ($w_{2}=0$) $N=2$ models: the values of $f$ so obtained were not unreasonable in comparison with experiments (17).

It should, perhaps, be mentioned in passing that Ref. (22) (see also (13)) invokes an Einstein relation in an analysis of ``protein friction". However, this is a rather different context in which many ``blocked" motor proteins are attached to a substrate and a rigid microtubule diffuses, apparently freely, close-by in the medium above. Quantitative arguments (13,\hspace{0.4mm}22) explain the large frictional slowdown observed (relative to an Einstein-relation estimate) as due to weak protein binding and unbinding.

{\bf Barometric Formulation.}\hspace{0.3em} Although the identification of the motor driving force $f$ with $f_{E}$ is unjustified, it would be desirable to have a soundly based, general expression for $f$ which, like $f_{E}$, does not entail any intrinsic modifications or extensions of the motor model or of the associated  physicochemical picture.  To that end, consider the placement of an ``impassable block" or {\it barrier} on the molecular track, say, between sites $L$ and $L+1$ ($\gg 1$) or at distance $x=D=l_{0}d$ from the origin $x=0$ (fixed, as we  suppose, by where the motor starts). Such a barrier may be realized  theoretically by decreeing that all states $j_{l}$ for $l \ge L+1$ are inaccessible; this may be achieved simply by setting one of the local forward rate constants, say, $u_{J+1}(l=L)$, equal to zero. No other rate constants need be modified; but if, perhaps to take cognizance of some aspects of a realistic barrier, further nearby rate constants are changed, it will have no consequences for the main conclusions.

It is intuitively clear that running a molecular motor up to such a barrier will lead (provided it does not detach from the track or ``freeze" irreversibly, as might happen in practice  (6,10)) to a stationary probability distribution, $P_{j}(l,t \rightarrow$$\infty)=P_{j}^{\infty}(L-l)$, in which $z=(L-l)d=D-x$ measures the distance back from the barrier. On very general theoretical grounds one should expect this distribution to decay exponentially with increasing $z$ (except for possible deviations close to the barrier) so that, explicitly, one has
\begin{equation}
P_{j}^{\infty}(z/d) \approx A_{j} e^{-\kappa z}.
\end{equation}
The (positive) decay constant $\kappa$ should, in principle, be experimentally measurable (although this may be difficult if $\kappa d$ is large). The amplitude ratios $A_{j}/A_{0}$ must depend on the various rate ratios, $u_{i}/w_{i}$, while $A_{0}$ is set simply by normalization.

To justify this surmise for the kinetic scheme in Eq. {\bf 1} (although it is of general validity), note that the mean flow between adjacent states $0_{l}$ and $(N-1)_{l-1}$ and between $j_{l}$ and $(j-1)_{l}$ [for $j=1,2, \dots,(N-1)$] must vanish for a stationary distribution. Balancing local forward and backward rates thus yields
\begin{eqnarray}
u_{N}(l-1) P_{N-1}^{\infty}(L-l+1)&=&w_{N}(l-1) P_{0}^{\infty}(L-l), \nonumber \\
\nonumber \\
u_{j}(l) P_{j-1}^{\infty}(L-l) &=& w_{j}(l) P_{j}^{\infty}(L-l),
\end{eqnarray}
for $j=1,2, \ldots ,(N-1)$. Starting from an initial nonzero value $P_{J}^{\infty}(0)$, one can then recursively determine $P_{J-1}^{\infty}(0),P_{J-2}^{\infty}(0)$, $\cdots, P_{0}^{\infty}(0),P_{N-1}^{\infty}(1),P_{N-2}^{\infty}(1),\cdots \hspace{2mm}$. By induction, this leads directly to Eq. {\bf 9} [since the $u_{j}(l)$, and $w_{j}(l)$ become independent of $l$ for, say, $l<L-l_{0}$ where $l_{0}$ is a fixed integer]. Most crucially one finds (with the notation of Eq. {\bf 2}) that the decay constant is simply
\begin{equation}
\kappa=(\ln{\Gamma})/d=\mbox{\boldmath $\varepsilon$}/d.
\end{equation}

Now, to interpret these results, consider a dilute gas of molecules of mass $m$ moving in a gravitational field that acts ``downwards" along the $z$ axis. Each molecule then has a weight $f_{G}=mg$; in addition, the equilibrium density distribution is given by (23)
\begin{equation}
\rho(z)=\rho(0) e^{-mgz/k_{B}T},
\end{equation}
where $\rho(0)$ is the density at the level $z=0$. (Any deviations arising close to the ``lower" wall (at $z \simeq 0$) due to molecular size, structure, etc., have been neglected.) Comparing this well known barometric formula with the distribution Eq. {\bf 9} leads us to identify the driving force $f$ of the molecular motor with
\begin{equation}
f_{B}=k_{B}T (\ln{\Gamma})/d=k_{B}T \mbox{\boldmath $\varepsilon$} /d.
\end{equation}
The subscript $B$ here serves merely to indicate the barometric analogy underlying our identification. By comparison with Eq. {\bf 6} for $f_{max}$, we may expect $\mbox{\boldmath $\varepsilon$}$  \raisebox{-0.6ex} {$\stackrel{<}{\mbox {\scriptsize ${\sim}$}}$}   $\Delta G_{0}/k_{B}T$ for a real molecular motor.

Before studying this result in relation to extensions of Eq. {\bf 1} needed to describe a motor functioning under external loads, let us compare $f_{B}$ with $f_{E}$.

{\bf Barometric vs. Einstein Scale.} \hspace{0.3em} Suppose the molecular motor operates close to equilibrium in the sense that $\mbox{\boldmath $\varepsilon$}=\ln{\Gamma}$ is small. Then, on expanding in $\mbox{\boldmath $\varepsilon$}$ at fixed $\omega/\sigma$, Eqs. {\bf 2}-{\bf 5} and {\bf 13} yield
\begin{equation}
f_{B}/f_{E}=1+[{\textstyle{\frac{1}{12}}}-(\omega/\sigma)]\mbox{\boldmath $ \varepsilon$}^2 - {\textstyle{\frac{1}{2}}} (\omega/\sigma)\mbox{\boldmath $ \varepsilon$}^{3} + \hspace{1mm} \cdots \hspace{1mm},
\end{equation}
for $N=2$. Evidently, the coefficient of $\mbox{\boldmath $\varepsilon$}$ vanishes identically! Furthermore, one finds $0 < \omega/\sigma \leq \frac{1}{16}$ so that the coefficient of $\mbox{\boldmath $\varepsilon$} ^2$ is small, lying between ${\textstyle{\frac{1}{48}}}$ and ${\textstyle{\frac{1}{12}}}$. Consequently, and as anticipated, the Einstein scale approximates the barometric result very well when the motor operates sufficiently close to equilibrium. Indeed, for $\Gamma <10$,  calculations show that $f_{B}$ can exceed $f_{E}$ by no more than $44 \%$. Furthermore, the series truncated at $O(\mbox{\boldmath $\varepsilon$}^2)$  in Eq. {\bf 14} is reasonably accurate up to $\mbox{\boldmath $\varepsilon$} \simeq 5$ ($\Gamma \simeq 150$) where one has $1.473 < f_{B}/f_{E} < 2.535$; beyond that, the bounds  $\frac{1}{4}\mbox{\boldmath $ \varepsilon$} < f_{B}/f_{E}$  \raisebox{-0.6ex} {$\stackrel{<}{\mbox {\scriptsize ${\sim}$}}$}  $\frac{1}{2} \mbox{\boldmath $\varepsilon$}$ are effective.

These  specific results are limited to $N=2$; but we suspect (and have checked for $N=3$) that the vanishing of the $O(\mbox{\boldmath $ \varepsilon$})$ term in Eq. {\bf 14} is independent of $N$.  Likewise, we expect $f_{B}$ always to rise steadily above $f_{E}$ when $\mbox{\boldmath $\varepsilon$}$ increases. Indeed, on recalling Eq. {\bf 2} for $\Gamma$, one observes from Eq. {\bf 13} that $f_{B}$ is unbounded above and so, with an injudicious assignment of rate constants, it may even exceed $f_{max}$ (as given in Eq. {\bf 1})!  Conversely, one may show from Eqs. {\bf 3, 4} and {\bf 8}, that $f_{E}$ for $N=2$ is bounded above by $4k_{B}T/d$ (17). However, we will demonstrate below that this bound on $f_{E}$ is rather artificial and does not apply for models that account directly for the discreteness of ATP hydrolysis.

{\bf Stalling Force Measured by Spring Compression.} \hspace{0.3em} In a typical  experiment on motor proteins (6\hspace{0.4mm}-10), optical tweezers are used to carry  a silica bead coated with a few  molecules of the motor protein up to the molecular track.  Then a single motor  binds to the track  and  starts to move, exerting a force against the  opposing load, $F$, as it pulls  the bead towards a side of the optical trap. The external force $F$ is a linear function of the displacement from the trap center, and the constant of proportionality can be measured. Thus the trap and bead  work like a calibrated spring acting on the molecular motor. To represent such an experiment the load-free scheme embodied in Eq. {\bf 1} must, clearly, be extended.

To this end, suppose the motor moves in a slowly varying external potential, $\Phi(x)$, so that in translocating from site $l$ to $l+1$, additional mechanical work  $\Delta \Phi(x=ld) =   \Phi(x+d)$   $ -\Phi(x)$ must be done (relative to the load-free situation). Of course, this corresponds to imposition of a local external force, $F(x)=\Delta \Phi(x)/d$, directed negatively. For an (ideal) optical trap of spring constant $K$ we may take
\begin{equation}
\Phi(x)={\textstyle{\frac{1}{2}}}  K x^{2}, \hspace{8mm} F(x) = K(x+{\textstyle{\frac{1}{2}}}d).
\end{equation}

In such a situation the motor should, in effect, compress the spring and, as $t$ increases, attain a stationary distribution, say $P_{0}^{S}(l)$, where, for simplicity, we focus only on the (free) states $0_{l}$. This distribution should peak at some $l_{S}$, corresponding to a mean (or most probable) compression of the spring by a  displacement $x_{S}=l_{S}d$. Then the measured ``stalling force" would be $f_{S}=Kx_{S}$.

Now it is physically clear that under any local load, $F(x)$, the transition rates, $u_{j}(l)$ and $w_{j}(l)$, must change. If, as traditional, one views the chemical transitions between successive states, $j$ and $j+1$, as proceeding in quasiequilibrium over various free energy barriers (13), one expects (in leading approximation) the rates to change exponentially with $F(x)d/k_{B} T$. But how the exponential loading factors should be distributed among the various reaction processes, $j \rightleftharpoons (j+1)$, is far from clear: indeed, this distribution is of considerable interest in understanding the motor mechanism at a microscopic level. Without prejudice, therefore, we will explore the {\it quasiequilibrium  hypothesis} that under a local load, $F$, the local transition rates change in accord with
\begin{eqnarray}
u_{j} \Rightarrow& u_{j}^{(F)}= &u_{j}^{(0)} e^{-\theta_{j}^{+} Fd/k_{B}T}, \nonumber \\
w_{j} \Rightarrow& w_{j}^{(F)}= &w_{j}^{(0)} e^{+\theta_{j}^{-}Fd/k_{B}T}.
\end{eqnarray}
The distribution factors, $\theta_{j}^{+}$ and $\theta_{j}^{-}$, need not be of uniform sign: but we certainly expect the overall factor
\begin{equation}
\theta= \sum_{j=1}^{N}(\theta_{j}^{+}+\theta_{j}^{-}),
\end{equation}
to be positive, implying an opposition to motion. Indeed, should the motor undergo diffusion in thermal equilibrium when {\it not} activated by ATP (as suggested parenthetically in the introductory discussion of Eq. {\bf 1}), detailed balance considerations would dictate $\theta =1$. As a {\it supplement} to our quasiequilibrium hypothesis this value of $\theta$ is also  plausible  for an activated motor that operates not too far from equilibrium. Notice  that a {\it negative} $\theta_{J}^{+}$ or $\theta_{J}^{-}$ simply means that the corresponding forward rate, $u_{J}$, is {\it enhanced}, or the reverse rate, $w_{J}$, is {\it diminished} by the internal molecular strain induced in the motor by the load.

Accepting Eq. {\bf 16} we can find the stationary distribution $P_{0}^{S}(l)$ with the aid of the rate-balance Eqs. {\bf 10} [ replacing $P_{j}^{\infty}(L-l)$ by $P_{j}^{S}(l)$ and the rates $u_{j}$ and $w_{j}$ in accord with Eq. {\bf 16} ].  The most probable motor location, $l_{S}$, follows by equating $P_{0}^{S}(l)$ and  $P_{0}^{S}(l+1)$ which leads directly to the condition
\begin{equation}
\Gamma^{(F)}(l) \equiv \prod_{j=1}^{N}[u_{j}^{(F)}(l)/w_{j}^{(F)}(l)]=\Gamma^{(0)} e^{-\theta F(x)d/k_{B}T}=1.
\end{equation}
Solving this determines $x_{S}=l_{S}d$ and thence yields the measured spring or stalling force
\begin{equation}
f_{S}= k_{B} T(\ln{\Gamma})/\theta d= k_{B} T \mbox{\boldmath $\varepsilon$}/ \theta d,
\end{equation}
where we have, of course, identified the zero-load rate factor, $\Gamma^{(0)}$, with the original rate factor $\Gamma$ in Eq. {\bf 2}.

It is striking that this expression for the stalling force (which depends on the quasiequilibrium hypothesis, Eq. {\bf 16}, that is needed to extend the original kinetic model) agrees {\it precisely} with the barometric expression Eq. {\bf 13} for $f_{B}$, provided one accepts the natural, near-equilibrium evaluation $\theta=1$. We regard this overall consistency as strengthening both approaches.

{\bf Velocity versus Load.} \hspace{0.3em} The extended rate constants $u_{j}^{(F)}$ and 
$w_{j}^{(F)}$ given in Eq. {\bf 16} also serve to provide a relation for the motor velocity, $V(F)$, as a function of a steady load force, $F$ [and, equally, for the load-dependent diffusion constant, $D(F)$]. For arbitrary $N$ one may appeal to Eq. {\bf A1} which shows, as expected, that the stalling load, $F_{S}$, which brings $V(F)$ to zero, agrees with Eq. {\bf 19}, i.e., $F_{S}=f_{S}$. To write an explicit result for $N=2$ in an illuminating form, we introduce the reduced force and modified load factors
\begin{equation}
\eta=F/F_{S} \hspace{8mm} {\rm {and}}  \hspace{8mm} \Delta_{j}^{\pm}={\textstyle {\frac{1}{2}}}-(\theta_{j}^{\pm}/\theta).
\end{equation}
Then from Eqs. {\bf 3, 16} and {\bf 19} we obtain
\begin{equation}
\frac{V(F)}{V(0)}=\frac{\sigma  \ {\rm {sinh}} [ {\textstyle {\frac{1}{2}}}\mbox{\boldmath $ \varepsilon$} (1-\eta)]/ {\rm {sinh}} ({\textstyle  {\frac{1}{2}}}\mbox{\boldmath $ \varepsilon$})}{ u_{1} e^{-\Delta_{2}^{+} \mbox{\scriptsize {\boldmath $\varepsilon$}} \eta} + u_{2} e^{-\Delta_{1}^{+}\mbox{\scriptsize{\boldmath $\varepsilon$}} \eta} + w_{1} e^{\Delta_{2}^{-}\mbox{\scriptsize{\boldmath $\varepsilon$}} \eta} + w_{2} e^{\Delta_{1}^{-}\mbox{\scriptsize{\boldmath $\varepsilon$}} \eta}},
\end{equation}
where, naturally, $V(0)$ is simply the no-load result of Eq. {\bf 3}; thus the right hand side reduces to unity when $\eta=0$ (and vanishes as $\eta \rightarrow 1$).

Now for $\mbox{\boldmath $\varepsilon$}$ small (say, \raisebox{-0.6ex}  {$\stackrel{<}{\mbox {\scriptsize ${\sim}$}}$}  $2$), so that the motor is operating not too far from equilibrium, one has $V(F) \approx V(0) (1-\eta)/$ $(1+c \mbox{\boldmath $\varepsilon$} \eta)$. This represents a {\it hyperbolic} force law which will be {\it concave} or {\it convex} depending on the sign, $+$ or $-$, of $c$ : see the illustrative examples in Fig. 1. For  small $c$ the law is close to {\it linear} and, in fact, $c$ vanishes whenever $u_{1} \Delta_{2}^{+} + u_{2} \Delta_{1}^{+} = w_{1} \Delta_{2}^{-} + w_{2} \Delta_{1}^{-}$. This condition has many solutions; for example, if the backward rates are small, so that $\delta=(w_{1}+w_{2})/(u_{1}+u_{2})$ \raisebox{-0.6ex}{$\stackrel{<}{\mbox {\scriptsize ${\sim}$}}$}   $0.1$, say, the loading scheme  $\theta_{1}^{+} \simeq  \theta_{2}^{+} \approx {\textstyle {\frac{1}{2}}} \theta/(1+\delta)$ yields a near-vanishing $c$.

On the other hand, if $u_{1}$ greatly exceeds $u_{2}$, $w_{1}$, and $w_{2}$, the reduced ($V,F$) plots become insensitive to $u_{1}$. Then if, as mentioned (6), one has $u_{1} \simeq k_{1}$[ATP], the plots will become independent of the ATP concentration (6). Furthermore, if $\mbox{\boldmath $ \varepsilon$}$ is large but $(\theta_{2}^{+}/\theta) \mbox{\boldmath $ \varepsilon$} \simeq 1$, the ($V,F$) plots will be close to linear.

Although straight, convex, and concave velocity-load plots are readily generated, other reasonable values of the six parameters: $\mbox{\boldmath $\varepsilon$}$, $w_{1}/u_{1}$,  $w_{2}/u_{1}$, and $\theta_{2}^{+}/\theta$, $\theta_{1}^{-}/\theta$ and $\theta_{2}^{-}/\theta$, yield plots exhibiting points of inflection of either sense, as shown in Fig. 1.  However, plots with {\it negative} inflection points, such as (e), are  realized  in relatively small regions of the parameter space.  If negative $\theta_{2}^{+}$ or $\theta_{1}^{+}$ are admitted (see after Eq. {\bf 17}) the velocity may initially {\it rise} when a load is imposed! Plots with two inflection points are then also allowed.  Thus  if   one could determine plausible values for the no-load transition-rate ratios, experimental $(V,F)$ plots might, at least for an $N=2$ model, throw some light on the load distribution parameters, $\theta_{j}^{\pm}$.

\begin{figure}[ht]
\centering
\centerline {\psfig {file= 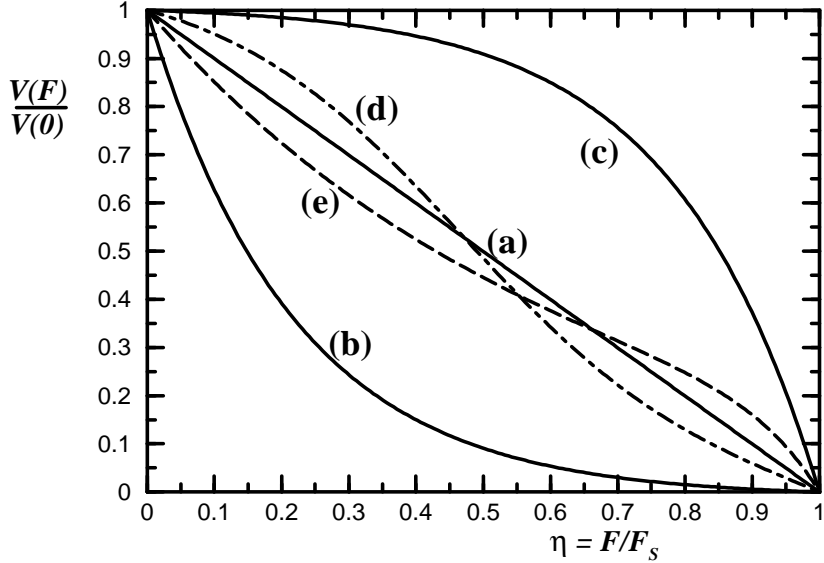,width= 12cm}}
\caption  {Examples of velocity-load plots for $N=2$ models with various parameter sets \{$\mbox{\boldmath{$\varepsilon$}}; (w_{1},w_{2})/u_{1}; (\theta_{2}^{+},\theta_{1}^{-}=\theta_{2}^{-})/\theta$\}: (a) \{$10^{-2}; 0.99,0.99;\frac{1}{2},0$\}, \newline
(b) \{$9.2; 10^{-2},10^{-2};\frac{1}{2},0$\}, (c) \{$10^{-2}; 10^{-2},10^{-2}; 0,\frac{1}{2}$\}, (d) \{$11.1; 10^{-4},0.15; 0,\frac{1}{2}$\}, \newline 
(e) \{$23.0; 10^{-5},10^{-5}; 0.07,0.43$\}.  Note that $V(0)$ is the velocity at zero load (Eq. {\bf 3}) while $F_{S}$ denotes the stalling load.}
\end{figure}

{\bf Discrete Jump Models.} \hspace{0.3em} As mentioned above,  the Einstein force scale obtained from the kinetic scheme, Eq. {\bf 1}, is subject to a fairly stringent bound. Analyzing the expression $f_{E}/k_{B}T=V/D$ (see Eq. {\bf 8}) for the case $N=2$ (using Eqs. {\bf 3}-{\bf 6}) we can prove that for all rates $u_{j}, w_{j} >0$, the Einstein scale satisfies $f_{E}/k_{B}T \leq 2N/d$; the maximum for $N=2$ is realized for uniform rates $u_{j}=u_{0} \gg w_{j}=w_{0}$ \hspace{-0.8mm}(all $j$), and we believe the same condition yields the bound as stated generally for all $N$. (The uniformity condition can be understood heuristically since in such a case there are no distinguishing rate-limiting steps in the cycle. The $N=2$ model studied in (17) also respects the {\it lower} bound $f_{E}/k_{B}T > 2/d$; but this is attributable to the special limiting situation, $w_{2}=0$, studied there which cannot be literally true.)

Our purpose here is to demonstrate that these bounds on $f_{E}$ are related to the continuous-time picture of the rate process embodied in the kinetic master equations based on Eq. {\bf 1}; in essence, these force a minimum value of the diffusion constant $D$. To see this most directly, consider an ($N=1$)-state model with master equation
\begin{equation}
\frac{\partial P_{0}}{\partial t}(l,t)=u P_{0}(l-1,t)+w P_{0}(l+1,t)-(u+w) P_{0}(l,t),
\end{equation}
where we have put $u_{1}=u \geq w_{1}=w > 0$. Then one finds
\begin{equation}
V=(u-w)d, \hspace{8mm} D={\textstyle{\frac{1}{2}}}(u+w) d^{2},
\end{equation}
[see, e.g. (16)]. Note the lower limit $D > \frac{1}{2} ud^{2}$, which is approached when $w/u \rightarrow 0$.  This leads directly to the bound $f_{E}/k_{B}T < 2/d$.

By contrast, consider a {\it discrete} event sequence in which a forward or backward jump is attempted at (mean) time intervals $\Delta t= \tau$ (triggered, one might picture for a molecular motor, by the arrival of individual ATP molecules). If $\check{P}_{0}(l;n)$ is the probability that the (motor) particle is at site $l$ after $n$ jump attempts, one has (19,20,24)
\begin{equation}
\check{P}_{0}(l;n+1)=p_{+} \check{P}_{0}(l-1;n)+p_{0} \check{P}_{0}(l;n)+p_{-} \check{P}_{0}(l+1;n),
\end{equation}
where $p_{+}$ and $p_{-}$ are the probabilities of completing a positive or negative step while $p_{0}=1-p_{+}-p_{-}$ is the probability of remaining at the same site. If one sets $p_{+}=u \tau$ and  $p_{-}=w \tau$, and identifies the time as $t \approx n \tau$, this discrete master equation, reduces to the continuous form, Eq. {\bf 22}, in the limit $\tau \rightarrow 0$ (24).

Now the mean displacement $\langle x \rangle_{n}$ after $n=1$ attempts is clearly $(p_{+}-p_{-})d$ so that the mean velocity is 
\begin{equation}
V=(p_{+}-p_{-})d/\tau=(u-w)d.
\end{equation}
Note that the identifications appropriate to the continuous limit yield agreement with Eq. {\bf 23}. To compute $D$ we may use Eq. {\bf 7} with only a {\it short} time interval, specifically $t=\tau$, since, by assumption, successive jump attempts are uncorrelated. Thus, from $\langle x^{2} \rangle_{1}=(p_{+}+p_{-})d^2$ we obtain
\begin{eqnarray}
D&=&{\textstyle{\frac{1}{2}}}(d^{2}/\tau)[ p_{+}+p_{-}-(p_{+}-p_{-})^{2}] \nonumber \\
 &=&{\textstyle{\frac{1}{2}}}[u+w-(u-w)^{2} \tau]d^{2}.
\end{eqnarray}

To see that $D$ now has no positive lower bound, we may specialize to the case $p_{0}=0$ or consider the limit $p_{-}(=w\tau) \ll p_{+}$: then one finds $D \propto (d^{2}/\tau) p_{+}(1-p_{+})$ which becomes indefinitely small when $p_{+}$ approaches unity. Hence no upper bound on $f_{E}$ exists in such a discrete jump model. Indeed, it is intuitively clear that in the limit $p_{+}=1$ (so that $p_{0}=p_{-}=0$) the particle moves essentially ballistically at speed $d/\tau$ with no dispersion.

Notice also that the barometric formulation can be applied directly to the jump model by using Eq. {\bf 24}. It leads precisely to the previous form, Eq. {\bf 13}, but with $\Gamma=p_{+}/p_{-}$; however, this agrees exactly with the  continuous-time ($N=1$) result $\Gamma=u/w$ when, as above, one puts $p_{+}=u \tau$ and $p_{-}=w \tau$. In addition, the ratio $R(\mbox{\boldmath $\varepsilon$})=f_{B}/f_{E}$ obeys Eq. {\bf 14} but with, in leading order, ($\omega/\sigma$) replaced by $\frac{1}{2} p_{-}=(1-p_{0})/2(\Gamma+1)$. For $\mbox{\boldmath $\varepsilon$} > 2$ one has $\frac{1}{2} p_{-} <0.06$ and $R(\mbox{\boldmath $\varepsilon$})$ varies much as discussed above for the continuous case.

It might be objected that  our arguments have more or less assumed that the jump attempts occur regularly at times $n \tau$ whereas, more realistically, there should be some  distribution, say  $\psi(t)$, of waiting times between one event and the next. Then $\tau$ would be the mean time between attempts, defined by
\begin{equation}
\tau=\overline {t}\hspace{4mm} \mbox { with } \hspace{4mm} \overline {t^{n}} =\int_{0}^{\infty} t^{n}\psi(t) dt.
\end{equation}      \renewcommand{\thefootnote}{\fnsymbol{footnote}}
Such a model may be studied along the lines of Montroll and Scher (24). Provided $\psi(t)$ decreases sufficiently fast  when $t \rightarrow \infty$ that the second moment $\overline{t^{2}}$ is finite, the analysis for $V$ and $D$ can be carried through: it shows again that  $D$ is unbounded below while $f_{E} \propto V/D$ is unbounded above. Indeed, Eq. {\bf 25} for $V$ remains valid while Eq. {\bf 26} for $D$ gains a factor $(1-\Theta)$ before each squared term, $(p_{+}-p_{-})^{2}$ and $(u-w)^{2}$. The parameter $\Theta=(\overline{t^{2}}-\overline{t}^{2})/\overline{t}^{2} \ge 0$ measures the relative width or ``spread'' of the waiting time distribution $\psi(t)$: e.g., for $\psi(t) \propto t^{\nu -1} e^{-\gamma t}$ with $\nu, \gamma >0$, one has\footnote{The specific results quoted in Ref. (24), Eqs. {\bf 75} for $\nu=\frac{1}{2}$ and $\nu= 2$ are in error (and the factor 4 in Eq. {\bf 76} should read 2).}  $\tau=\nu/\gamma$ and $\Theta=1/\nu$. The sharp distribution originally envisaged corresponds to the limit $\nu \rightarrow \infty$.

Finally, note that we can also analyze precisely  multistate versions of these discrete jump models with waiting times.

\section*{\hspace{20mm} \bf DISCUSSION AND SUMMARY}

In order to understand the driving force, $f$, exerted by a molecular motor that takes steps of size $d$ on a molecular track, we have analyzed a broad class of stochastic models: in particular, Eqs. {\bf 1} and {\bf 2} embody a general, ``linear'' reaction sequence. In the presence of a constant free energy source, the motor will  achieve a steady velocity $V$ ($>0$) but with  fluctuations about the mean position, $\langle x(t) \rangle =Vt$,  described by a diffusion constant, $D$, that  may be measured by observing the variance: see Eq. {\bf 7}. The table below lists various force scales that then arise and their relation to $f$.

\begin{table}[ht]
\caption{{\bf Forces related to  a Molecular Motor}}
\vspace{5mm}
\centering
\begin{tabular}{|llr|} \hline
Maximum driving force  & $f_{max}\hspace{1.5mm} =\Delta G_{0}/d > f$  & [Eq. {\bf 6}]  \\
Einstein Scale         & $\hspace{5mm}f_{E}  =k_{B}TV/D < f $         & [Eq. {\bf 8}]  \\ 
Gravitational force    & $\hspace{5mm}f_{G}  =mg$                     & [Eq. {\bf 12}] \\ 
Barometric force       & $\hspace{5mm}f_{B}  =k_{B}T\kappa \cong f$   & [Eqs. {\bf 12},{\bf 13}]  \\
Stalling  force  &$\hspace{5mm}f_{S} =k_{B}T\mbox{\boldmath $\varepsilon$}/\theta d \stackrel{?}{=}f$& [Eq. {\bf 19}]  \\ 
Load and stalling load& $F$, $F_{S}$,\hspace{3mm} $\eta=F/F_{S}$ &[Eq. {\bf 20}]  \\ \hline
\end{tabular}
\end{table}

By way of a concrete numerical illustration,  consider a kinesin molecule moving on a microtubule (2\hspace{0.4mm}-4, 6\hspace{0.4mm}-10, 12) for which $d \simeq 8.2$ nm (9). Svoboda, Mitra, and Block (6) observed  $V \simeq 670$ nm/s when [ATP] $=2$ mM and  measured the variance from which we obtain $D \simeq 1395$ nm$^{2}$/s. At $T=300$ K these data yield $f_{E} \simeq 2.0$ pN. On the other hand, the observed stalling force was $f_{S} \simeq 5$-$6$ pN (6, 7) significantly larger than $f_{E}$,  as we have argued it should be. Note also, comparing with the  maximal force, $f_{max} \simeq 6.2$ pN, that the observed efficiency $\mbox{\LARGE {$\varepsilon$}}$, is in the range 80-95{\%}. (This observational estimate does {\it not} allow for the possible ``wastage'' of ATP by futile hydrolysis (13) without translocation of the motor: recall the footnote below Eq. {\bf 5}.)

The  ``barometric'' force scale, $f_{B}$,  arises by considering an obstacle that blocks the motor's motion  on the track: the resulting  statistically stationary  distribution  decays with the distance $z$ from the obstacle  as  $e^{-\kappa z}$: see Eq. {\bf 9}.  It would be interesting (although difficult)  to measure $\kappa$  and to compare $f_{B}$, so derived, with the observed stalling force $f_{S}$.

For the general ($N=2$)-state model (see Eqs. {\bf 1}-{\bf 5}) with transition rates $u_{1}$, $u_{2}$, $w_{1}$ and $w_{2}$  one has  $f_{B}=(k_{B}T/d) \times \ln(u_{1} u_{2}/w_{1} w_{2})$. For kinesin (from {\it Drosophila}) Gilbert and Johnson (12) studied the kinetics using chemical-quench flow methods. Assuming [ATP] $= 2$ mM  their data show that  $u_{1}=3800$ s$^{-1}$, $u_{2}=15$ s$^{-1}$, and $w_{1}=200$ s$^{-1}$ represents  a sensible map on to an $N=2$ model; however, $w_{2}$ proved unobservably small. Merely for illustration, therefore,  suppose $w_{2}=u_{2}/100=0.15$ s$^{-1}$. This gives  $V \simeq 116$ nm/s and $D \simeq 474$ nm$^{2}$/s, which yield  $f_{E} \simeq 1.0$ pN (at $T=300$ K),  while the rates give  $f_{B} \simeq 3.8$ pN. The  agreement with the results of Svoboda {\it et al.} is not impressive: nevertheless,  the orders of magnitude, the inequality $f_{B} > f_{E}$, and the rough equality $f_{B} \simeq f_{S}$, are in full accord with our analysis. 

More recently, Higuchi {\it et al} (9) obtained  data (for bovine brain kinesin) leading us to $u_{1} \simeq 1400$ s$^{-1}$ and  $u_{2} \simeq 45$ s$^{-1}$, in only rough  agreement with the values derived from (12). The {\it ad hoc} assumption  $w_{1}/u_{1} \simeq w_{2}/u_{2} \simeq 1/100$  yields $V \simeq 354$ nm/s and $D \simeq 1370$ nm$^{2}$/s, closer to  observations (6). Likewise,   $f_{E} \simeq 1.1$ pN and $f_{B} \simeq 4.7$ pN, now accord better  with the direct experiments (although depending logarithmically on  $w_{1}$ and $w_{2}$). While the general theoretical picture is supported, further  experiments on standardized kinesin samples would clearly be valuable and could  provide more stringent tests.

To discuss the velocity  $V(F)$ of  a motor  under a load $F$, the  transition rates must be modified:  see Eqs. {\bf 16} and {\bf 17} where the  load-distribution factors,  $\theta_{j}^{\pm}$, recognize that the various transitions in a motor protein probably accept quite different fractions of the total stress. Indeed, some forward rates might even be {\it accelerated} which could provide a mechanism to conserve, e.g., ATP under ``no-load'' conditions.  It is natural to take the overall load-distribution factor $\theta$ (Eq. {\bf 17}) as unity which leads to the equality of $f_{S}$ and $f_{B}$: see Eq. {\bf 19} {\it et seq.}   However, $\theta=1$ can be doubted  for real  motors  and might well be  tested by experiment or simulation.

Even for a two-state model, the expression for $V(F)$  is quite complex: see Eq. {\bf 21}. As seen in Fig. 1, the six independent parameters  permit  velocity-load plots of varied shapes (including nonmonotonic forms not shown). Certain types, such as  (e) characterize small regions of the  parameter space; but, in   general,  the variation of $V$ with $F$ may reveal comparatively  little about the motor mechanism or parameter values. 

Negative, i.e., assisting loads ($F <0$) are predicted to speed up  the motor and  this  has  been observed (10). Conversely, under super-stalling loads ($F>F_{S}$), {\it backwards velocities} are predicted;  single reverse steps of kinesin have then  been seen  (10) but no  steady reverse velocities have  been reported. This probably reflects  very small terminal reverse  rates, $w_{N}$  (12). Indeed, these transitions  presumably  describe  second (or higher) order chemical reactions  controlled by the  low  concentrations of hydrolysis products.   The frequently observed   process of  detachment  from the  track (6,10)  should also be included  in a fuller account.

The adequacy of the  stochastic models encompassed by  Eq. {\bf 1} is challenged  by  lower bounds on the diffusion constant, $D$, which  yield upper bounds on  $f_{E}$. For kinesin at $T=300$ K this bound is 2.03 pN for any ($N=2$)-state model.  The data of Svoboda {\it et al.} (6,7)  essentially meet this; but were the  bound violated, one might  conclude that an $N\ge3$ kinetic  model was needed.  However, models in which the transitions are described by {\it discrete jumps} occurring  after certain  waiting times, are  {\it not} susceptible to these constraints.  Such models might well prove more realistic, although at present the simpler kinetic representations seem  adequate. Nevertheless,  it should be noted that the main principles we have  enunciated are not restricted to the $N=2$  sequential kinetic models specifically analyzed. Consequently, the observation of significant violations would indicate serious deficiencies in the general understanding of  motor mechanisms.

\section*{ }
We are indebted to   David A. Huse, Stanislas Leibler, Michelle D. Wang and Benjamin Widom for valuable  comments  on our work. The support of the National Science Foundation (under Grant CHE 96-14495) is gratefully acknowledged.

\appendix
\section*{Appendix}

\renewcommand{\theequation}{A\arabic{equation}}

\setcounter{equation}{0}

For a one-dimensional hopping model with $N$ states and arbitrary  transition rates $u_{j}$  and $w_{j}$, as introduced in Eq. {\bf 1},  Derrida (16) obtained the exact steady-state behavior. For the drift velocity he found 
\begin{equation}
V=\frac{d}{\sum_{j=1}^{N}r_{j}} \left(1-\prod_{j=0}^{N-1} \frac{w_{j}}{u_{j}} \right ), 
\end{equation} 
where $d$ is the spatial period (or step-size) while
\begin{equation}
r_{j}=\frac{1}{u_{j}}\left(1+\sum_{k=1}^{N-1} \prod_{i=1}^{k} \frac{w_{j+i-1}}{u_{j+i}}\right).
\end{equation}
The expression for the diffusion constant (16) is similar but more complex and less illuminating.

For  $N=2$,  solutions can be obtained for all times following the procedure outlined in (17). Thus, the probability that the particle is at site  $l$ in state $j_{l}$ ($j=0,1$) after a time $t$ having started at the origin, $l=0$, is
\begin{equation}
P_{j}(l,t) = \int_{-\pi}^{\pi} \frac{dq}{2\pi}  e^{-iq(l+j/2)}  \left[  \Xi_{+}(q) e^{\lambda_{+}(q)t}- \Xi_{-}(q) e^{\lambda_{-}(q)t} \right ] 
\end{equation}
with rate parameters $\lambda_{+}(q)$ and $\lambda_{-}(q)$  given by 
\begin{equation}
\lambda_{\pm}(q)={\textstyle{\frac{1}{2}}}\left[-\sigma \pm \sqrt{\sigma^2+4u_{1}u_{2}(e^{2iq}-1)+4w_{1}w_{2}(e^{-2iq}-1)} \right], 
\end{equation}
while $\sigma$ is  in Eq. {\bf 5} and  the coefficient functions  are 

\begin{equation}
\Xi_{\pm}(q)=  \frac{\lambda_{\mp}(q)+u_{1}+w_{2}}{\lambda_{-}(q)-\lambda_{+}(q)} \left (1+\frac{\lambda_{\pm}(q)+u_{1}+w_{2}}{u_{2}e^{iq}+w_{1}e^{-iq}} \right). 
\end{equation}



\end{document}